\documentclass[12pt,preprint]{aastex}

\slugcomment{\it Accepted to the Astrophysical Journal Letters}
\lefthead{Mahabal et al.}
\righthead{An Optically-faint Quasar at $z=5.70$}

\def\c1788{RD~J114816.2+525339}
\def\rd1788{RD~J1148+5253}

\def\cf{{cf.,~}}

\def\eg{{e.g.,~}}
\def\etal{{et al.}}

\def\deg{\ifmmode {^{\circ}}\else {$^\circ$}\fi}
\def\kms{\ifmmode {\rm\,km\,s^{-1}}\else
    ${\rm\,km\,s^{-1}}$\fi}
\def\ergcm2s{\ifmmode {\rm\,ergs\,cm^{-2}\,s^{-1}}\else
    ${\rm\,ergs\,cm^{-2}\,s^{-1}}$\fi}
\def\ergAcm2s{\ifmmode {\rm\,ergs\,cm^{-2}\,s^{-1}\,\AA^{-1}}\else
    ${\rm\,ergs\,cm^{-2}\,s^{-1}\,\AA^{-1}}$\fi}
\def\ergs{\ifmmode {\rm\,ergs\,s^{-1}}\else
    ${\rm\,ergs\,s^{-1}}$\fi}
\def\kmsMpc{\ifmmode {\rm\,km\,s^{-1}\,Mpc^{-1}}\else
    ${\rm\,km\,s^{-1}\,Mpc^{-1}}$\fi}

\def\lya{Ly$\alpha$}
\def\lyb{Ly$\beta$}
\def\nv{\ion{N}{5} $\lambda$1240}

\def\spose#1{\hbox to 0pt{#1\hss}}
\def\simlt{\mathrel{\spose{\lower 3pt\hbox{$\mathchar"218$}}
     \raise 2.0pt\hbox{$\mathchar"13C$}}}
\def\simgt{\mathrel{\spose{\lower 3pt\hbox{$\mathchar"218$}}
     \raise 2.0pt\hbox{$\mathchar"13E$}}}

\begin{document}

\title{Discovery of an Optically-Faint Quasar at $z=5.70$ and \\
Implications for the Faint End of the Quasar Luminosity Function}

\author{A.~Mahabal\altaffilmark{1},
D.~Stern\altaffilmark{2},
M.~Bogosavljevi\'c\altaffilmark{1},
S.G.~Djorgovski\altaffilmark{1},
\& D.~Thompson\altaffilmark{1}}

\altaffiltext{1}{Palomar Observatory 105--24, California Institute of
Technology, Pasadena, CA 91125; aam@astro.caltech.edu, milan@astro.caltech.edu,
george@astro.caltech.edu, djt@irastro.caltech.edu}

\altaffiltext{2}{Jet Propulsion Laboratory, California Institute of
Technology, Mail Stop 169-506, Pasadena, CA 91109; stern@thisvi.jpl.nasa.gov}


\begin{abstract}

We present observations of an optically-faint quasar, \c1788 (hereafter
\rd1788), discovered from deep multi-color observations of the field
around the $z = 6.42$ quasar SDSS~J1148+5251.  The two quasars have a
projected separation of 109\arcsec\ and both are outliers in $r - z$
versus $z - J$ color-color space.  Keck spectroscopy reveals \rd1788\
to be a broad-absorption line quasar at $z = 5.70$.  With $z_{\rm AB}
= 23.0$, \rd1788 is 3.3 mag fainter than SDSS~J1148+5251, making it the
faintest quasar known at $z>5.5$.  This object was identified in a survey
of $\approx 2.5$ square degrees.  The implied surface density of quasars
at these redshifts and luminosities is broadly consistent with previous
extrapolations of the faint end of the quasar luminosity function and
supports the idea that active galaxies provide only a minor component of
the reionizing ultraviolet flux at these redshifts.

\end{abstract}



\section{Introduction}

The increasing optical depth of the \lya\ forest in high-redshift
quasars provided the first evidence that the cosmic reionization epoch
concluded at $z \approx 6$, approximately 1~Gyr after the Big Bang
\markcite{Djorgovski:01, Becker:01}(Djorgovski {et~al.} 2001; Becker {et~al.} 2001).  While active galaxies no longer appear
to be significant contributors to the reionization of the intergalactic
medium \markcite{Yan:04}(\eg Yan \& Windhorst 2004), quasars still provide essential probes of
the ionization state of the universe, as well as probes of the earliest
stages of galaxy formation.  For example, the sizes of \ion{H}{2} regions
around the highest redshift quasars provide strong constraints on the
cosmic neutral fraction at early epoch \markcite{Wyithe:04, Mesinger:04}(Wyithe \& Loeb 2004; Mesinger \& Haiman 2004).
Probing the faint end of the quasar luminosity function is also
important for understanding the interplay between the formation
of galaxies and the formation of supermassive black holes.

The Sloan Digital Sky Survey (SDSS) has found many luminous ($M_{1450}
\simlt -26.5$) quasars at high ($z \simgt 5.5$) redshifts \markcite{Fan:01,
Fan:03}(Fan {et~al.} 2001, 2003) and determined the evolution of the bright end of the quasar
luminosity function \markcite{Fan:04}(Fan {et~al.} 2004).  Others, such as \markcite{Wolf:03}Wolf {et~al.} (2003)
and \markcite{Hunt:04}Hunt {et~al.} (2004), have studied the faint end of the quasar luminosity
function at lower redshifts ($z \simlt 3$).  In terms of faint quasars
at high redshifts ($z \simgt 5$), few sources have been reported.
\markcite{Djorgovski:03}Djorgovski {et~al.} (2003) reported a $z = 4.96$ quasar slightly fainter than
the SDSS limits ($z_{\rm AB} = 21.2$; $M_B = -25.2$) and within a few
Mpc of the $z = 5.02$ quasar SDSS~0338+0021.  \markcite{Stern:00c}Stern {et~al.} (2000)
identified a single $z = 5.50$ faint ($z_{\rm AB} = 23.4$; $M_B =
-22.7$) quasar in a small-area survey designed to find high-redshift
Lyman-break galaxies.  Interestingly, this source is well detected
at 1.2 mm by MAMBO, implying a far-infrared luminosity $L_{\rm FIR}
\approx 4 \times 10^{12} L_\odot$, comparable to the average luminosity
of high-redshift SDSS quasars which are an 3 magnitudes more luminous at
optical wavelengths \markcite{Bertoldi:02, Staguhn:05}(Bertoldi \& Cox 2002; Staguhn {et~al.} 2005).  \markcite{Barger:02}Barger {et~al.} (2002)
identified an X-ray selected $z = 5.189$ faint ($z_{\rm AB} = 23.7$)
quasar in the {\it Chandra} Deep Field - North.  Searches for additional
AGN at these high redshifts in the combined {\it Chandra} Deep Fields by
\markcite{Barger:03b}Barger {et~al.} (2003) and \markcite{Cristiani:04}Cristiani {et~al.} (2004) have yielded negative results.
Recently, there have been a few concerted observational programs to map
large areas of sky to faint magnitudes, with the goal of identifying
high-redshift, low-luminosity quasars: \markcite{Sharp:04}Sharp {et~al.} (2004) reports on $VIz$
mapping of 1.8~${\rm deg}^2$ of sky, going approximately 2~mag fainter
than the SDSS, while \markcite{Willott:05}Willott {et~al.} (2005) reports on $i'z'$ mapping of
3.8~${\rm deg}^2$ of sky, going approximately 3~mag fainter than the SDSS.
To date, neither survey has identified any new, high-redshift quasars.

In this {\it Letter} we report on early results of a wide-area, multiband
program to identify faint, high-redshift quasars.  Using the Palomar
and Keck observatories, we have obtained $riz$ images of approximately
2.5~${\rm deg}^2$, going approximately 3~mag fainter than the SDSS.
For portions of our survey, we have also obtained near-infrared imaging.
With the goal of identifying large-scale structure in the early universe
\markcite{Djorgovski:03, Stiavelli:05}(\cf Djorgovski {et~al.} 2003; Stiavelli {et~al.} 2005), while still being sensitive
to unassociated, high-redshift sources, we have primarily imaged
high-redshift SDSS quasar fields.  Detailed results from this study will
be presented elsewhere (Bogosavljevi\'c \etal, in preparation); here we
present observations of a faint quasar at $z=5.70$ found in the field
of the highest redshift quasar currently known, SDSS~J1148+5251 at $z =
6.42$ \markcite{Fan:03}(Fan {et~al.} 2003).  Throughout we adopt a $\Lambda$-cosmology with
$\Omega_{\rm M} = 1 - \Omega_\Lambda = 0.3$ and $H_0 = 65~ \kmsMpc$.
At $z=5.70$, such a universe is 1.05~Gyr old, the lookback time is
92.7\%\ of the total age of the universe, and an angular size of 1\farcs0
corresponds to 6.3~kpc.


\section{Observations}

We obtained images of the SDSS~J1148+5251 field in the $R$- and
$z$-bands using the Low Resolution Imaging Spectrometer
\markcite{Oke:95}(LRIS; Oke {et~al.} 1995) at the 10-meter Keck-I telescope on UT 2003
April 6, and in the $J$-band using the Wide-Field Infrared Camera
\markcite{Wilson:03}(WIRC; Wilson {et~al.} 2003) at the Palomar 200-inch Hale telescope on UT
2003 May 22.  The total exposure times were 300, 1000, and 3600 seconds
in the $R$-, $z$-, and $J$-bands, respectively.  LRIS has a 5\arcmin\
$\times$ 7\arcmin\ field-of-view, while WIRC has a 8.7\arcmin\ $\times$
8.7\arcmin\ field-of-view.


The images were reduced using standard procedures.  The seeing FWHM is
1\farcs2 in the processed optical images and 0\farcs9 in the processed
$J$-band image.  As conditions were not photometric for either imaging
nights, we estimated the calibrations by comparison with the SDSS (for the
optical data; AB magnitudes) and the Two Micron All Sky Survey (for the
near-infrared data; Vega magnitudes).  Due to a relatively large scatter
in the SDSS cross-matching, likely due to differences between the SDSS
and Keck/LRIS filter transmission functions, we estimate 0.3 magnitude
uncertainties in the optical photometry.  This level of photometric
accuracy is more than sufficient for our purposes.

We generated $R$ and $z$ catalogs using Source Extractor
\markcite{Bertin:96}(Bertin \& Arnouts 1996) in double-image mode, using the $z$-band image for
source detection.  The $J$-band photometry was performed separately,
using astrometrically-matched catalogs.  Optical magnitudes were
measured using 2\farcs5 diameter apertures and the near-IR magnitudes
were measured in 3\farcs5 diameter apertures.  Based on source counts, we
estimate the images reach the following depths:  $R_{\rm lim} \approx 27$,
$z_{\rm lim} \approx 25$, and $J_{\rm lim} \approx 22.5$.  In the $r -
z$ versus $z - J$ color-color diagram (Fig.~1), two objects satisfy the
criteria for high-redshift objects, $R - z \geq 3.5$.  One is the quasar
SDSS~J1148+5251 itself, and the other is \c1788 (hereafter \rd1788)
-- the object described herein.  Table~\ref{table.phot} presents the
imaging properties of these sources and Fig.~\ref{fig.finder} shows
their relative positions.

We obtained a spectrum of \rd1788\ on UT 2003 May 23 with LRIS on the
the Keck~I telescope under partly cirrusy conditions.  Observations
totaled 4800~sec and used the 400 line mm$^{-1}$ grating ($\lambda_{\rm
blaze} = 8500$~\AA) at a position angle of $117.5\deg$.  Spectral
reductions followed standard procedures.  The final spectrum is
presented in Fig.~\ref{fig.spectrum}.


\section{Results and Discussion}

The spectrum of \rd1788\ reveals a slightly atypical quasar at $z =
5.70$.  The main spectroscopic feature is the unambiguous detection
of highly redshifted \lya\ emission, showing the characteristic
asymmetric profile due to absorption of the blue wing of the emission
line \markcite{Stern:05a}(\cf Stern {et~al.} 2005). 
The quoted redshift is based on the
peak of the Ly$\alpha$ emission line.  This standard approach, the only
option for discovery optical spectra of extremely distant quasars,
typically overestimates the true redshift by $\Delta z \approx 0.05$ as
determined from near-infrared spectroscopic follow-up
(e.g., Goodrich et al. 2001, Barth et al. 2003, Stern et al. 2004).
The spectrum also shows broad
\nv\ emission (FWHM $\approx 2700~ {\rm
km s}^{-1}$) and evidence of spectral breaks associated with the
\lya\ and \lyb\ forests.  The continuum falls dramatically blueward
of \nv, suggesting that \rd1788\ is a broad absorption line quasar.
Such quasars constitute approximately 10\%\ of the quasar population,
and this self-absorption is perhaps somewhat responsible for the unusually
narrow \lya\ emission (FWHM $\approx
900~ {\rm km s}^{-1}$) of \rd1788.
Alternatively, the narrow \lya\ could be coming from a low-density,
intermediate-line region while the \nv\ could be coming from a
higher-density, very broad line region (c.f., Brotherton et al. 1994).

Carilli et al. (2004) have imaged the SDSS~J1148+5251 field down to several $\mu{\rm Jy}$ at 1.4~GHz.
Slightly offset from the optical position, about $1''$ to the E, 
they find a faint, $2.5\sigma$ detection of a source with a
peak flux of $44 \pm 18\ \mu{\rm Jy}$ (Carilli, private communication). From
available data, determining the emission mechanism isn't feasible
for such a faint, low significance detection.

How surprising is the discovery of the faint, high-redshift quasar \rd1788?
Similar programs reported by \markcite{Sharp:04}Sharp {et~al.} (2004) and \markcite{Willott:05}Willott {et~al.} (2005)
failed to identify any new, high-redshift, faint quasars, leading
\markcite{Willott:05}Willott {et~al.} (2005) to infer that the co-moving space density of
quasars brighter than $M_{1450} = -23.5$ declines by a factor $> 25$
from $z = 2$ to $z = 6$.  The $R$-band dropout criterion applied here
selects sources at $z \simgt 5$ \markcite{Stern:00c}(\cf Stern {et~al.} 2000).  Based on a
large sample of faint ($R < 24$) quasar candidates selected from the
COMBO-17 survey, \markcite{Wolf:03}Wolf {et~al.} (2003) derive the most recent, comprehensive
evaluation of the faint end of the quasar luminosity function out to $z
= 5$.  Extending the pure density evolution version of their luminosity
function to redshifts slightly beyond where it has been tested, we expect
a surface density of 0.72 (2.08) $z > 5$ quasars per square degree to a
$z$-band limiting magnitude $z_{\rm lim} = 23$ ($z_{\rm lim} = 24.5$).
The former magnitude limit corresponds to the brightness of \rd1788,
while the fainter limit corresponds to the approximate depth of our
$z$-band imaging for the SDSS~J1148+5251 field.  Considering only the
SDSS~J1148+5251 field which covers $\approx 1\%$ of a square degree,
the discovery of \rd1788\ would be quite fortuitous and would suggest
a significant evolution in the faint end of the quasar evolution at
high redshift.  However, considering the full 2.5 square degree survey
we have conducted, the discovery of a single $z \simgt 5$, $z_{\rm AB}
\approx 23$ quasar implies a surface density roughly consistent with the
\markcite{Wolf:03}Wolf {et~al.} (2003) luminosity function.  We conclude that previous estimates
of the faint, high-redshift quasar luminosity function based on surveys
with null results \markcite{Sharp:04, Willott:05}(\eg Sharp {et~al.} 2004; Willott {et~al.} 2005) and surveys at
slightly lower redshift \markcite{Wolf:03, Hunt:04}(\eg Wolf {et~al.} 2003; Hunt {et~al.} 2004) are broadly
correct:  active galactic nuclei make a negligible contribution to the
ultraviolet radiation budget and are unlikely to be play a significant
role in reionizing the universe at $z \approx 6$.


\acknowledgements

We thank Chris Carilli for providing the radio flux
from their deep observations of the quasar field, and an anonymous
referee for pointing out their work. We also thank Mike Brotherton
for useful comments.
The data presented herein were obtained at the W.M. Keck Observatory,
which is operated as a scientific partnership among the California
Institute of Technology, the University of California and the National
Aeronautics and Space Administration. The Observatory was made possible by
the generous financial support of the W.M. Keck Foundation.  The authors
wish to recognize and acknowledge the very significant cultural role
and reverence that the summit of Mauna Kea has always had within
the indigenous Hawaiian community; we are most fortunate to have the
opportunity to conduct observations from this mountain.  This work was
supported in part by the NSF grant AST-0407448 and by the Ajax Foundation.
The work of DS was carried out at Jet Propulsion Laboratory, California
Institute of Technology, under a contract with NASA.






\begin{deluxetable}{lcccccc}
\tablecaption{Imaging Results}
\tablehead{
\colhead{Object} &
\colhead{R.A.} &
\colhead{Dec.} &
\colhead{$R$} &
\colhead{$z$} &
\colhead{$J$} &
\colhead{$M_B$}}
\startdata
SDSS~J1148+5251 & 11:48:16.67 & +52:51:50.4 & $25.2\pm0.3$ & $19.7\pm0.3$ & $18.07\pm0.02$ & $-27.8$\\
\rd1788          & 11:48:16.21 & +52:53:39.3 & $> 27.0$     & $23.0\pm0.3$ & $21.45\pm0.06$ & $-24.3$\\
\enddata

\tablecomments{Astrometry is in J2000 coordinates.  Optical magnitudes
are in the AB system; near-IR magnitudes are in the Vega system.
$M_B$ is obtained by transforming the $z_{AB}$ magnitude to $B_{AB}$ using the
\markcite{VandenBerk:01}Vanden Berk {et~al.} (2001) quasar template spectrum.}

\label{table.phot}
\end{deluxetable}


\begin{figure}
\plotone{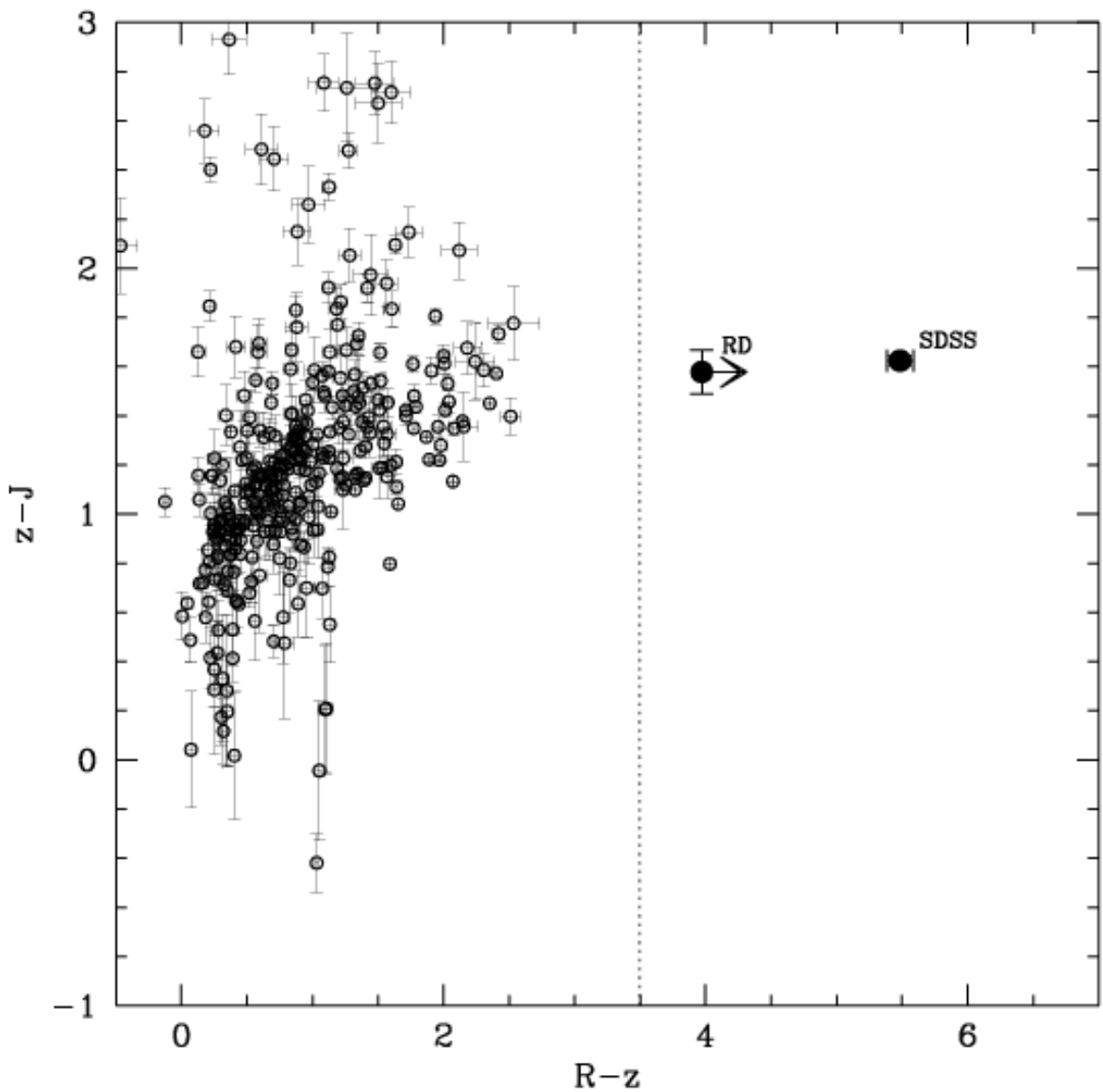}
\caption{Color-color diagram for the SDSS~J1148+5251 field.  Optical
magnitudes are in the AB system; near-IR magnitudes are in the Vega
system.  The vertical dotted line indicates the color criterion used to
identify high-redshift quasar candidates.  Error bars on the photometry
indicate only the formal statistical errors on the measurements,
as determined by Source Extractor.  Photometry presented in Table~1
includes the systematic uncertainty in the optical zeropoints.
Area included in this figure is $\sim 0.01$ square degrees.}
\label{fig.colcol}
\end{figure}


\begin{figure}
\plotone{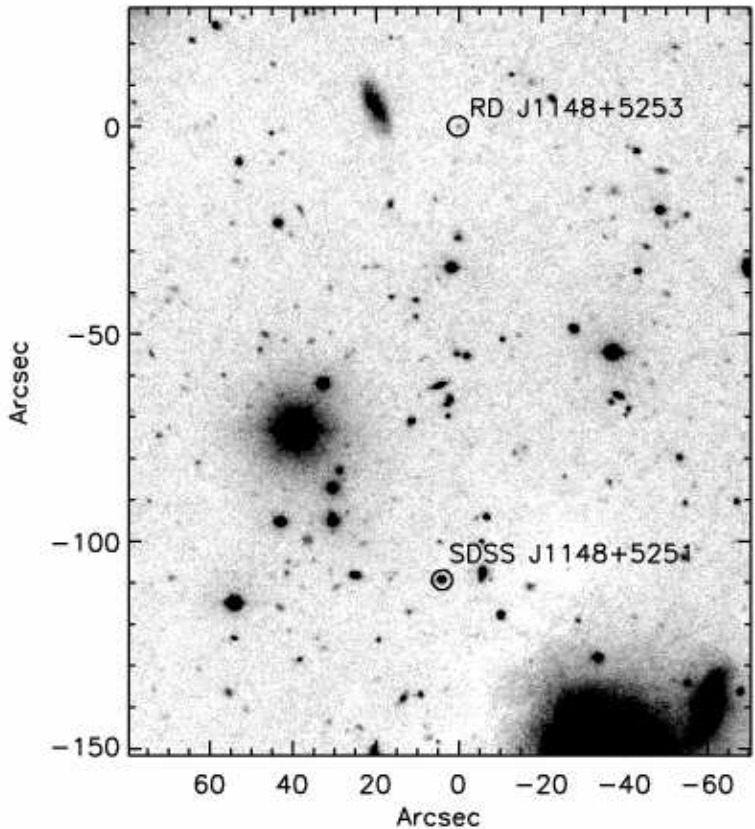}
\caption{Finding chart for the SDSS~J1148+5251 field, from the LRIS
$z$-band image.  The field size is 150\arcsec\ $\times$ 180\arcsec, with
north up and east to the left.  The two quasars, labeled, are separated
by 109\arcsec.}
\label{fig.finder}
\end{figure}


\begin{figure}
\epsscale{0.9}
\plotone{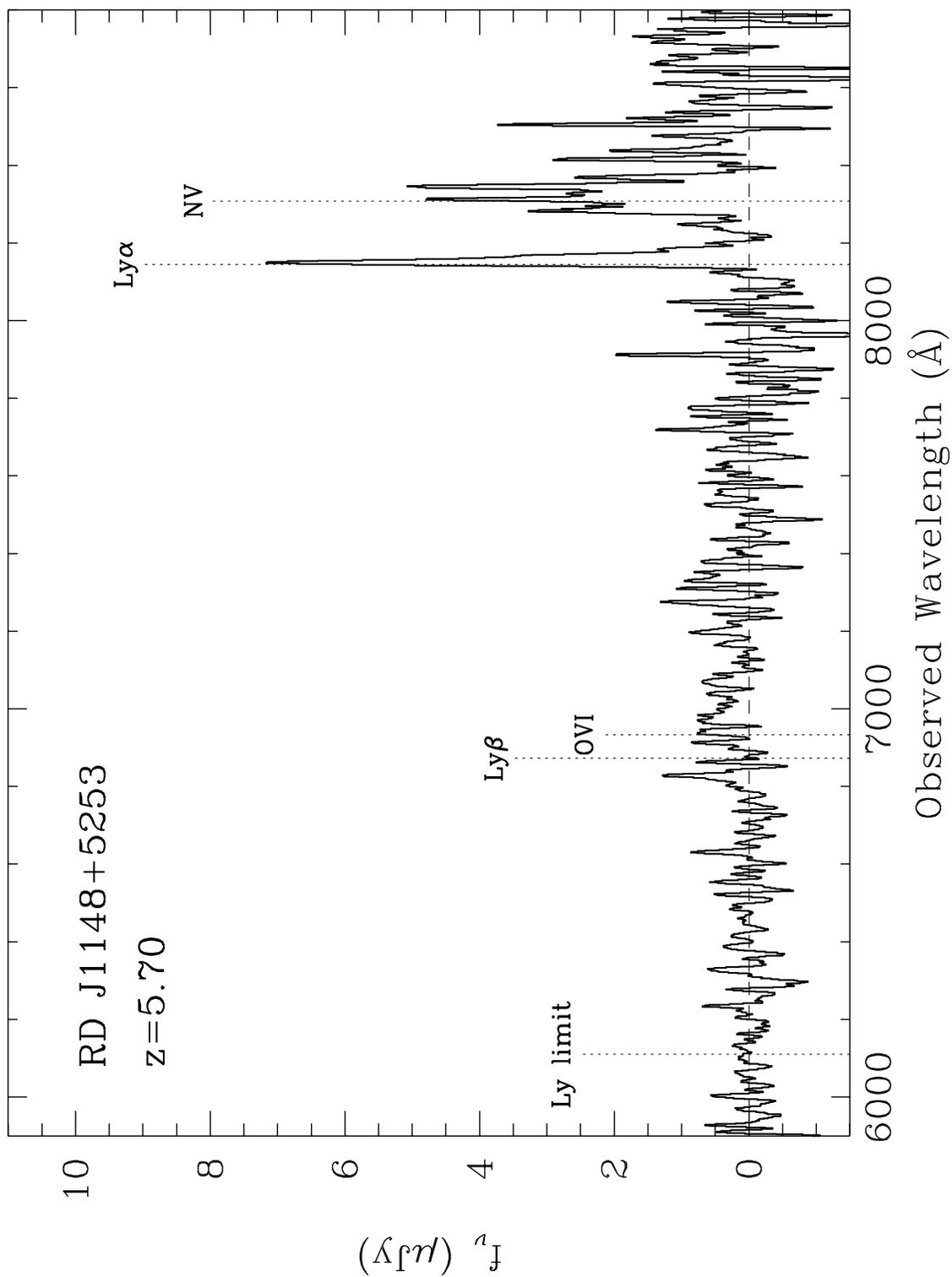}
\caption{Spectrum of \rd1788\ at $z = 5.70$, obtained with LRIS on the
Keck~I telescope.  Prominent features are indicated.  As observations
were obtained under non-photometric conditions, the flux calibration is
estimated.}
\label{fig.spectrum}
\end{figure}


\end{document}